\documentclass[a4paper,twocolumn]{revtex4}

\usepackage{graphicx,natbib}
\usepackage{mathrsfs}
\usepackage{times}
\usepackage{amsmath}

\newcommand{\br}{{\mbox{\boldmath$r$}}}
\newcommand{\bu}{{\mbox{\boldmath$u$}}}
\newcommand{\bK}{{\mbox{\boldmath$K$}}}
\newcommand{\los}{{\mbox{\boldmath$\hat{\ell}$}}}
\newcommand{\bk}{{\mbox{\boldmath$k$}}}
\newcommand{\bx}{{\mbox{\boldmath$x$}}}
\newcommand{\by}{{\mbox{\boldmath$y$}}}
\newcommand{\bv}{{\mbox{\boldmath$v$}}}
\newcommand{\F}{{\mathscr F}}
\newcommand{\G}{{\mathscr G}}

\begin{document}

\title{Sensitivity of solar f-mode travel times to internal flows}

\author{Jason~Jackiewicz}
\affiliation{Max Planck Institut f\"{u}r Sonnensystemforschung, Katlenburg-Lindau, 37191, Germany}
\author{Laurent Gizon}
\affiliation{Max Planck Institut f\"{u}r Sonnensystemforschung, Katlenburg-Lindau, 37191, Germany}
\author{Aaron C. Birch}
\affiliation{Colorado Research Associates, a Division of Northwest Research Associates Inc., Boulder, CO, 80301, USA}

\begin{abstract}
We compute f-mode travel-time sensitivity kernels for flows.  Using a two-dimensional model, we show that it is important to account for several systematic effects, such as the foreshortening and the projection of the velocity vector onto the line of sight. Correcting for these effects is necessary before any data inversion is attempted away from the center of the solar disk.
\end{abstract}
\maketitle
\section{Introduction}
The solar f-modes are useful tools for studying perturbations that reside
within the top $2$~Mm of the surface of the Sun, and have been used as a
diagnostic as such before, e.g., \citet{duvall2000}. The linear forward problem of
time-distance helioseismology is to calculate functions, or kernels, which
give the sensitivity of travel-time measurements to these particular
perturbations of the f-modes. For a discussion of this general procedure,
which is the one that we will follow  here, see \citet{gizon2002}. 

In this study we are interested in \textit{flow} kernels, i.e., functions
which give the two-dimensional sensitivity of travel-time measurements to small amplitude, two-dimensional, spatially varying flows. The flow kernel, $\bK$, is defined as
\begin{equation}
\delta\tau=\int\!\!\int_S d^2\br\,\bu(\br)\cdot\bK(\br),
\label{kern-def}
\end{equation}
where $\delta\tau$ is the perturbation to the travel time, $\bu=(u_x,u_y)$ is the two-dimensional flow perturbation, and the integral is over the position vector $\br=(x,y)$ over the whole solar surface. The kernel $\bK=(K_x,K_y)$ is a two-dimensional vector which has sensitivity to flows in the $\bx$ and $\by$ directions.

We calculate the kernels from an
approximate solar model, which utilizes the Born approximation in estimating
the scattering from solar inhomogeneities. This has been shown to be more accurate than calculating kernels from a ray approximation \citep{birch2004}, since it includes finite-wavelength effects.

Ultimately, the goal is to perform inversions with real time-distance
measurements to extract the unknown $\bu$ from Eq.(\ref{kern-def}), given
$\delta\tau$ and $\bK$.  However, there are several difficulties in
doing this since the travel-time measurements incur systematic effects from the
observing process, particularly due to the center-to-limb variations. To overcome these problems, one strategy is to
calculate the sensitivity kernels from the outset accounting for the
systematic effects so that the whole procedure - the forward calculation and the
inversions - remains consistent.  

We have accomplished this for two important effects: the foreshortening of the
data pixel, and the projection of the solar velocity field onto the line of
sight, both of which will be defined in detail below. 
We present an analysis of the contribution that these two
effects have on the power spectrum of f-modes and on the kernels. For
brevity, we leave the details of the full calculation of the power spectra and kernels for a
future publication \citep{jackiewicz2006}, and only focus here on the role of the
systematic effects mentioned above on the final results. 
\section{model power spectrum}
\subsection{Line of sight projection}
A local region of the Sun may be approximated by a plane which is tangent to
the sphere.  The orientation of the normal to the plane with
respect to the observer depends on the position on the sphere on which this
plane lies.  We denote the
line of sight unit vector, $\los=(\los_h,\ell_z)$, as the direction which
points towards the observer from the point where the plane intersects the sphere. In
this notation,
$\ell_z=1$ only at the disk center. Doppler measurements of oscillation velocities $\bv$ on the Sun return the projection of $\bv$ onto $\los$. 
\begin{figure*}[t!]
\centering
\includegraphics[width=17.5cm, height=7.5 cm]{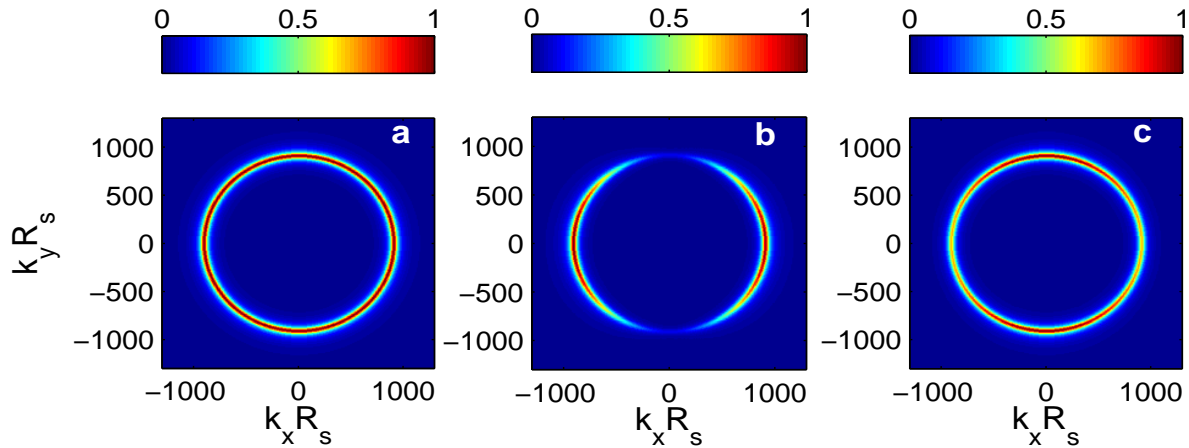}
\caption{A cut in three different power spectra at constant $\omega/2\pi=3$~mHz with $\Theta=90^o$, where the $x$ and $y$ axes are in units of harmonic degree. (a) shows the power at the disk center ($\Phi=0$) for reference. (b) is the power at $\Phi=75^o$, with the effect of the line of sight only (no foreshortening). (c) is again at  $\Phi=75^o$, but with the effect of foreshortening included (no line of sight effect).  All figures have been normalized by their maximum values, and $R_s=696 Mm$. We use a pixel size of an MDI hi-res image, $dx=0.43$~Mm (see section \ref{foreshortening}).}
\label{fig:3power}
\end{figure*}
For the purpose of inversions, in order to have confidence in utilizing all of
the data from instruments such as MDI and HMI, even data well away from disk center,
the sensitivity kernels in principle should account for a changing  $\los$, as to
avoid the aforementioned inherent systematic errors. In the past, kernels have been
calculated only with $\ell_z=1$ as a first attempt. We calculate kernels here for any $\los$.

In general, the observable oscillation signal, $\psi(\br,t)$, can be written as
\begin{equation}
\psi(\br,t)=\F\left\{\los(\br)\cdot\bv(\br,t)\right\}.
\end{equation}
The operator $\F$ involves the filtering procedure of the data, such as the
time duration of the measurement, intrumental effects such as the MTF and the
foreshortening, and possibly some phase speed filter. Using a similar model
for f-modes as in \citet{gizon2002}, and similar notation, we calculate an
expression for $\bv(\br,t)$. The zero-order power spectrum of $\psi$ (in the absence of a flow) can then be written for horizontal wavenumber $\bk$ and angular frequency $\omega$ as
\begin{equation}
P(\bk,\omega)=(2\pi)^6|\G(\bk,\omega)|^2,
\label{power}
\end{equation}
where
\begin{equation}
\G(\bk,\omega)=\left(i\los_h\cdot\bk+\ell_z\kappa(\omega)\right)\sqrt{m_s(\omega)}G(k,\omega)F(\bk,\omega).
\label{curlyg}
\end{equation}
Here, $\kappa$ describes the frequency dependent damping, $m_s$ is the source
power, $G$ is the Green's function, and $F$ is a function that describes the effect of the filter operator $\F$ mentioned above. For the purposes here, the important thing
to note is the explicit dependence of the power on the line of sight
vector, which is given as
$\los=(-\sin\Phi,-\cos\Theta\cos\Phi,\sin\Theta\cos\Phi)$, where $\Phi$, the solar longitude, is
measured from the central meridian, and $\Theta$ is the co-latitude measured from the solar north pole. In our local plane coordinates, $\Phi$ is along $\hat{\bx}$, and $\Theta$ is along $-\hat{\by}$.

Consider the power along the equator ($\Phi$ varies, $\Theta$ fixed at
$90^o$), and without any foreshortening effects for now.  Upon close inspection of
Eq.(\ref{curlyg}), one sees that the magnitude of the power along the $k_x$ direction is
mostly unaffected, while the power along $k_y$ is reduced. This is shown in Fig.~\ref{fig:3power}, where we plot cuts at constant angular frequency in the power spectrum. The first panel, Fig.~\ref{fig:3power}a, is at disk center, a reference plot where $\los=\los_z$. Fig.~\ref{fig:3power}b is a power cut at $\Phi=75^o$ towards the western limb, on the equator.  As stated above, the power is reduced along the $k_y$ direction. 

\begin{figure}[b]
\centering
\includegraphics[width=7.5 cm, height=5.5 cm]{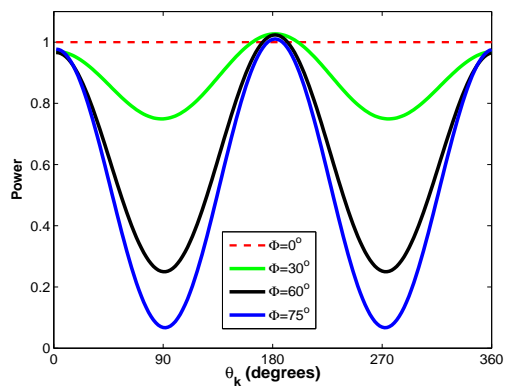}
\caption{Power in normalized units versus $\theta_k$ for the 4 diffent values of $\Phi$ along the equator ($\Theta=90^o$), given in the legend.  Here, $\ell=900$ and
  $\omega/2\pi=3$~mHz, corresponding to the power maximum of the f-mode. Note how only along $k_y$ ($\theta_k=90^o$,$\theta_k=270^o$) does the power get reduced due to the line of sight projection. Each curve is divided by the power at disk center (red dashed line) for normalization.}
\label{fig:los_v_thetak}
\end{figure}

To see this effect even more clearly, Fig.~\ref{fig:los_v_thetak} is a plot of a cut in the  power spectrum at constant $\omega$ for 4 different
positions on the disk at fixed $|\bk|$, versus the direction of $\hat{\bk}$ (denoted by the angle $\theta_k$ that $\bk$ makes with the $\bx$ axis), neglecting for the moment foreshortening effects. One sees that the power is always reduced along $k_y$
with respect to the value at the disk center.  The power along $k_x$ is
independent of $\Phi$ when only line of sight effects are considered. 

Physically, we can understand this by considering the extreme case when we observe at the
limb, on the equator.  Waves that travel north-south will be undectable because the motion is perpendicular to the line of sight, while
waves that travel along the equatorial plane still give power. 

\subsection{Foreshortening}
\label{foreshortening}
Foreshortening occurs due to the fact that the spatial resolution (measured on the Sun) decreases as we observe toward the limb. After projection onto the local plane, a ccd pixel will then appear elongated in the center-to-limb direction. Again, consider a region towards the limb on the equator.  A pixel that has a
horizontal resolution of $dx$ at disk center, now images a horizontal
distance on the Sun of $dx/\mu$, where $\mu=\cos\Phi\sin\Theta$. This reduces the effective spatial resolution measured on the Sun in the center-to-limb direction, and hence leads to
systematic observational errors.

\begin{figure}[t]
\centering
\includegraphics[width=8 cm, height=5.5 cm]{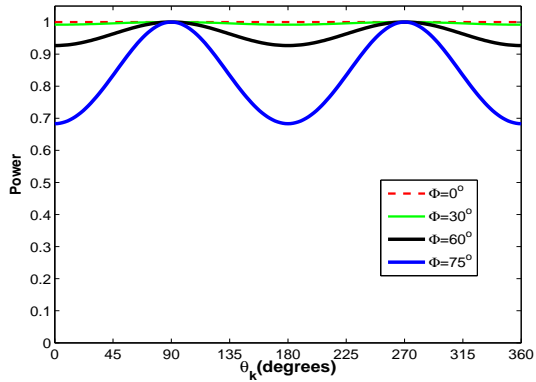}
\caption{Same as in Fig.~\ref{fig:los_v_thetak} with all curves at $\Theta=90^o$, but now we show the power taking into account the foreshortening $\Lambda(\bk)$ without any line of sight effects. Note how the foreshortening is $\pi/2$ out of phase with the line of sight (compare with Fig.~\ref{fig:los_v_thetak}).}
\label{fig:fore_v_thetak}
\end{figure}
\begin{figure*}[t]
\centering
\includegraphics[width=18 cm, height=7.5 cm]{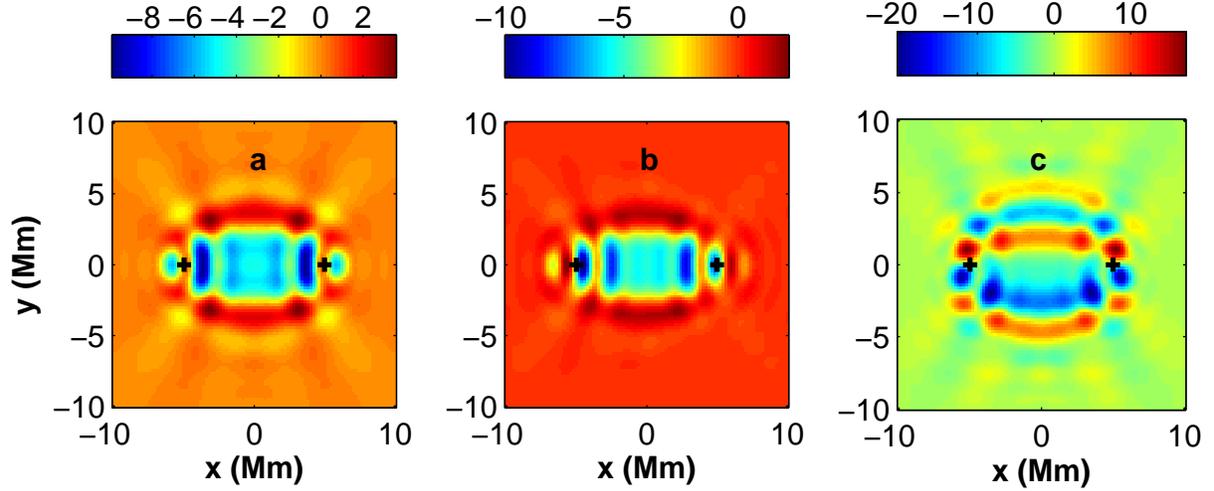}
\caption{F-mode sensitivity kernels $K_x$ for flows in the $+\hat{\bx}$ direction. (a) is calculated at disk center ($\Phi=0$, $\Theta=90^o$), (b) for $\Phi=75^o,\Theta=90^o$, and (c) for $\Phi=0, \Theta=15^o$. The crosses at $\br_1=(-5,0)$~Mm, $\br_2=(5,0)$~Mm denote the two observation points. The units of the colorscale are $s(km/s)^{-1}Mm^{-2}$. Note that the integral over space of each kernel is approximately the same.}
\label{fig:kerns}
\end{figure*}

To correct for this, we return to the expressions for the power given above in
Eqs.(\ref{power},\ref{curlyg}). The filter function $F(\bk,\omega)$ contains the information
about the foreshortening. We
break the filter into its constituent parts, and rewrite it as 
\begin{equation}
F(\bk,\omega)=f(k,\omega)\Lambda(\bk),
\end{equation}
where the lowercase $f(k,\omega)$ contains the MTF and/or other filters, which only depend on $k$, and $\Lambda(\bk)$ is the foreshortening correction. For the sake of simplicity, we obtain $\Lambda(\bk)$ by taking the Fourier transform of an elliptical disk, which is approximately the projection of a ccd pixel onto the local plane. We require that the ellipse only gets elongated
along the center-to-limb direction, which is
always along the semi-major axis.  The semi-minor axis stays at fixed length. Carrying
out this integration gives a term proportional to a Bessel function of first
order.  Expanding the Bessel function in a small parameter $k$ (wavenumber)
results in an expression given as
\begin{equation}
\label{fore_ell}
\Lambda(\bk)\approx \Lambda_0(k) + \Lambda_1(k)\sin^2(\theta_k-\alpha),
\end{equation}

where $\Lambda_0(k)$ and $\Lambda_1(k)$ are only functions of $k$ and are of
little interest here, except to note that they depend on the eccentricity of the ellipse, and
hence on $\Theta$ and $\Phi$.  We denote by $\alpha$ the angle which describes the orientation of the ellipse on the disk. More precisely, it is the angle that the semi-major
axis of the ellipse makes with a local latitude line, and is also a function of $\Theta$ and $\Phi$. All of the
other details of this calculation are beyond the scope of this paper, and will be presented elsewhere.

The foreshortening term (\ref{fore_ell}) now can be inserted into the power to study its
effects.  Returning to Fig.~\ref{fig:3power},  in panel (c) we plot the power including the foreshortening function, but this time neglecting the line of sight
effect. We see that the foreshortening effect is 90 degrees out
of phase with the line of sight effect from panel Fig.~\ref{fig:3power}b. The power is reduced instead
along the $k_x$ direction.  We can understand this physically, because waves are attenuated at high wavenumber, due to the smaller pixel size in Fourier space in the center-to-limb direction.

As was done for the line of sight case, we show a cut of the power at $\ell=900$, $\omega/2\pi=3$~mHz versus $\theta_k$ in Fig.~\ref{fig:fore_v_thetak} to see the reduced power in the $k_x$ direction more clearly as the limb is approached.  This figure, along with Fig.~\ref{fig:los_v_thetak}, demonstrates that the net reduction of power is always larger due to line of sight effects, i.e., at $\Phi=75^o$, the power is only about $10$~\% of its disk center value along $k_y$, while it is still about $70$~\% of its disk center value along $k_x$. The evidence for this is also in Fig.~\ref{fig:3power}b, where the power is almost gone along $k_y$, but is still somewhat present in panel (c) along $k_x$.  

\section{kernels}
We calculate the flow kernels $\bK$ based on the general method described in \citet{gizon2002}, where kernels were calculated only for damping and source perturbations to f-modes.  The flow kernels here are computed in a completely analagous manner: we use the single-scattering Born approximation. The effects of the line of sight and foreshortening are explicitly included in the calculation.

Since we are interested in detecting flows, we consider travel-time \textit{differences} between two points on the Sun. Because waves that travel in the direction of a flow travel faster than they would against it, this definition of travel time difference is sensible for extracting flows.  In Fig.~\ref{fig:kerns} we plot kernels for travel-time differences that are sensitive to flows travelling in the $+\hat{\bx}$ direction ($K_x$), where the two observation points are given by the left and right crosses. Fig.~\ref{fig:kerns}a is computed for the center of the grid at disk center. For flow perturbations between the observation points (blue region), one sees a negative value of the kernel, as expected.  In other areas, wave effects produce non-trivial behavior.  
Fig.~\ref{fig:kerns}b, computed well towards the western limb, demonstrates how the line of sight and foreshortening effects alter the kernel, as does Fig.~\ref{fig:kerns}c, which is a plot of a kernel on the central meridian at $\Theta=15^o$. The kernel in Fig.~\ref{fig:kerns}c shows quite different structure, in fact, it has very little sensitivity along the line connecting the two observation points; only waves that scatter off of the unperturbed ray path can be detected. The decomposition of the sensitivity kernel into Fresnel zones (e.g. \citet{jensen2001}) is not obvious in this last example. Also, note that even though the integral over space of the three kernels in Fig.~\ref{fig:kerns} is approximately the same, the spatial-dependence is quite different, and an inversion using these kernels would give different results for flows varying on a scale comparable to or smaller than the wavelength of f-modes.

\section{conlusions}
We have calculated sensitivity kernels for flows for time-distance helioseismology.
The effects of the line of sight and the foreshortening, as defined above, have been demonstrated for a model power spectrum and the kernels. The next step is to study real data for f modes to compare the observed power spectra with the modeled ones to gain confidence in our theoretical sensitivity kernels.

\end{document}